\documentclass[preprint,showpacs,preprintnumbers,amsmath,amssymb]{revtex4}
\usepackage{dcolumn}
\usepackage{bm}
\usepackage{graphicx}

\begin{document}


\title{Closure of the single fluid magnetohydrodynamic equations \\ in presence of electron cyclotron current drive}

\author{E. Westerhof}
\email[E-mail: ]{E.Westerhof@differ.nl}
\author{J. Pratt}
\affiliation{ FOM Institute DIFFER, Dutch Institute for Fundamental Energy Research,  \\
Nieuwegein, Netherlands, www.differ.nl}

\date\today

\begin{abstract}
  \noindent In the presence of electron cyclotron current drive (ECCD), the Ohm's law of single fluid magnetohydrodynamics (MHD) is modified as ${\bf E} + {\bf v} \times {\bf B} = \eta( {\bf J} - {\bf J}_{\rm EC} )$. This paper presents a new closure relation for the EC driven current density appearing in this modified Ohm's law. The new relation faithfully represents the nonlocal character of the EC driven current and its main origin in the Fisch-Boozer effect. The closure relation is validated on both an analytical solution of an approximated Fokker-Planck equation as well as on full bounce-averaged, quasi-linear Fokker-Planck code simulations of ECCD inside rotating magnetic islands. The new model contains the model put forward by Giruzzi et al., {\it Nucl. Fusion {\bf 39} (1999) 107}, in one of its limits.
\end{abstract}


\maketitle

\section{Introduction}

The evolution of large-scale instabilities in magnetized plasmas is commonly studied in the framework of fluid models like single fluid magnetohydrodynamics (MHD)~\cite{goedbloed}. These fluid models are obtained by taking moments of the governing kinetic equations. This results in a hierarchy of equations in which the evolution of each of the moments depends on higher order moments. At some point this moment expansion is truncated by introducing closure relations that model higher order moments in terms of lower order moments. In some cases, MHD instabilities can be affected by particular kinetic processes that determine important aspects of the velocity space distribution of one or more particle species. Such cases demand either a hybrid kinetic-fluid description or a closure of the fluid equations that takes into account these specific kinetic processes. An example is the case of electron cyclotron current drive (ECCD), which has become the primary tool in present-day tokamak operation for the control of MHD instabilities \cite{prater2004}. On the time scale of the macroscopic plasma evolution, the effect of electron cyclotron heating and current drive is described in the kinetic equation by quasi-linear diffusion of the electron distribution in velocity space. Because of the high parallel velocity of resonant electrons, electron cyclotron heating and current drive results in a highly nonlocal modification of the distribution function. In RF current drive modelling this is dealt with by bounce averaging of the kinetic equation \cite{killeen}, which assumes that the equilibration over a flux surface is achieved sufficiently quickly, so that the RF driven velocity space perturbation and current are effectively constant over a flux surface. Moreover, the velocity-space perturbation is established on a collisional time scale, which in high temperature tokamak plasmas can be of the same order as the time scale for the evolution of MHD instabilities. A closure of the MHD equations in the presence of ECCD should reflect all these aspects.

In a recent paper~\cite{hegna2009}, Hegna and Callen discuss the general framework for the fluid closure in the presence of RF heating and current drive. They show that the RF quasi-linear diffusion directly enters the single fluid MHD equations as a power source in the energy balance, and as a parallel force in Ohm's law. In addition it affects the closure for the resistivity through its effect on the electron-ion friction. In the case of ECCD, the quasi-linear diffusion is dominantly in the direction of perpendicular momentum~\cite{prater2004}, and the ECCD parallel force contribution in Ohm's law is negligible. Instead, the EC current drive is effected mostly by the creation of an asymmetric collisionality, known as the Fisch-Boozer effect \cite{fischboozer}. This effect must be contained in Ohm's law entirely through the proper closure relation for the resistivity, i.e. the electron-ion friction. Assuming a linear response of the plasma to the driving forces, the total current density ${\bf J}$ becomes the sum of the inductively driven current, $(1/\eta) ({\bf E} + {\bf v} \times {\bf B})$, with the usual (neoclassical) Spitzer resistivity $\eta$, and the non-inductive current from ECCD, ${\bf J}_{\rm EC}$: i.e. Ohm's law becomes
\begin{equation}\label{ohm}
    {\bf E} + {\bf v} \times {\bf B} = \eta( {\bf J} - {\bf J}_{\rm EC} ).
\end{equation}
This is common practice for modeling the effects of ECCD and other non-inductively driven currents on MHD. The subtraction of the non-inductive currents in Ohm's law also forms the basis for the generalized Rutherford equation, which describes the effects of noninductive current perturbations like ECCD on the evolution of neoclassical tearing modes (NTMs)~\cite{hegna1997,zohm1997,lahaye2006}. To model ECCD stabilization of tearing modes, this modification Ohm's law~(\ref{ohm}) is also used in three dimensional, numerical MHD simulations~\cite{yu2000,fevrier2014}. The EC driven current is evolved from a separate kinetic calculation or evolved simultaneously using a closure model. An ad hoc model for the description of ${\bf J}_{\rm EC}$ has been proposed by Giruzzi et al. in~\cite{giruzzi1999}.

In this work a new closure model for the evolution of ${\bf J}_{\rm EC}$ is derived through approximation of the governing kinetic equation. The resulting model faithfully represents the nonlocal character of the driven current and its origin in the Fisch-Boozer mechanism~\cite{fischboozer}. It contains the model of Giruzzi et al. as one of its limits. This work is structured as follows. Section~2 presents a short review of the most relevant aspects of ECCD. In particular, the quasi-linear RF diffusion operator appearing in the Fokker-Planck equation for the electron distribution function is discussed. Next, the quasi-linear Fokker-Planck equation is reduced to an analytically solvable equation through a series of approximations that are relevant for ECCD applied to a tokamak. The analytical solution is used to illustrate the highly nonlocal character of the EC current generation. These results are used to motivate our new closure model for the evolution of the EC driven current density. In Section~3, validation of the proposed closure model is provided. A first validation shows the excellent reproduction of the analytical solutions for the EC driven current density obtained with the approximated quasi-linear Fokker-Planck equation derived in Section~2. A second validation is performed by comparison of results from our new closure model with the evolution of the EC driven current density inside a rotating magnetic island as obtained from bounce-averaged quasi-linear Fokker-Planck calculations~\cite{ayten2014}. Again good agreement is obtained between our new closure for the EC driven current density and the Fokker-Planck calculations. The final section provides a brief summary and discussion of the results. An early report on this work was made at the $18^{\rm th}$ Joint Workshop on Electron Cyclotron Emission and Electron Cyclotron Resonance Heating~\cite{westerhof2014}.

\section{A new closure relation for the EC driven current density}

\subsection{Essential features of ECCD}

The kinetic description of ECCD is based on the Boltzmann equation. After averaging over the short-time scales of the electron gyromotion and the waves, the gyrophase-averaged Boltzmann equation is
\begin{equation}\label{boltzmann}
    {\partial f_{\rm e} \over \partial t} = C(f_{\rm e}) + Q_{\rm EC}(f_{\rm e}) - v_\parallel \nabla_\parallel f_{\rm e},
\end{equation}
where $f_{\rm e}(t, {\bf x}, v_\parallel, v_\perp )$ is the gyrophase-averaged electron velocity distribution as a function of parallel and perpendicular velocities, respectively, $C(f_{\rm e})$ represents the effect of collisions, quasi-linear diffusion, $Q_{\rm EC}(f_{\rm e})$, models the averaged effect of the electron cyclotron waves, and the final term describes the convection along magnetic field lines of localized features in the electron distribution function with their parallel velocity.

The effect of EC waves, with frequency $\omega$ and wave vector $k$, on the electron distribution function is well described by quasi-linear theory~\cite{prater2004}. For simplicity, we consider the non-relativistic limit in which the quasi-linear diffusion operator becomes \cite{kennelandengelmann,karney1986}:
\begin{equation}\label{qloperator}
    Q_{\rm EC}(f_{\rm e}) = {\partial \over \partial {\bf v}} \cdot {\bf D}_{\rm EC} \cdot {\partial \over \partial {\bf v}}
    f_{\rm e}
\end{equation}
\begin{equation}\label{DRF}
    {\bf D}_{\rm EC} = {\pi \over 2} {e^2 \over m_{\rm e}^2} \delta(\omega - k_\parallel v_\parallel - n \Omega_{\rm e}) {\bf
    a}_n^* {\bf a}_n
\end{equation}
with
\begin{equation}\label{an}
    {\bf a}_n = {1 \over \omega}
    \left( (\omega - k_\parallel v_\parallel) \hat {\bf v}_\perp + k_\parallel v_\perp \hat{\bf v}_\parallel \right)
    \left({\tilde E_- J_{n-1} + \tilde E_+ J_{n+1} \over \sqrt{2}} + {v_\parallel \over v_\perp} J_n \tilde E_\parallel \right)
\end{equation}
where $\hat{\bf v}_{\parallel,\perp}$ are unit vectors in the parallel and perpendicular velocity direction, respectively, $\Omega_{\rm e}$ is the electron cyclotron frequency, and $\tilde {\bf E}$ is the wave electric field with components of left and right handed circular polarization $\tilde E_{\pm}$, respectively, and parallel component $\tilde E_\parallel$. The $J_n$ represent the $n^{\rm th}$ order Bessel function of argument $k_\perp v_\perp / \Omega_{\rm e}$. In the case of electron cyclotron resonance, $\omega - k_\parallel v_\parallel = n \Omega_{\rm e}$, the ratio of parallel over perpendicular diffusion is
\begin{equation}
    {a_{n,\parallel} \over a_{n,\perp}} = {k_\parallel v_\perp \over n \Omega_{\rm e}} \ll 1.
\end{equation}
Because under typical experimental conditions the parallel refractive index for electron cyclotron waves is $|N_\parallel| < 1$, the quasi-linear diffusion due to electron cyclotron waves is dominantly in the perpendicular direction. In particular, for O-mode waves at fundamental resonance, which is the mode of choice for ECCD in large tokamaks like ITER~\cite{IPB_Chapter6,ramponi2008}, we can approximate the quasi-linear diffusion operator as
\begin{equation}~\label{approx_omode}
    {\bf D}_{\rm EC} \approx D \delta(v_\parallel - v_{\parallel,\rm res}) \hat v_\perp \hat v_\perp,
\end{equation}
where $v_{\parallel,\rm res} = (\omega - n \Omega_{\rm e})/ k_\parallel$ is the parallel velocity of the resonant electrons.

In spite of the near absence of direct momentum transfer between the electron cyclotron waves and the resonant electrons, a net current nevertheless is generated since the EC driven perturbation of the distribution function creates an asymmetric collisionality~\cite{fischboozer}. The generation of this `Fisch--Boozer' current can be understood as follows. Using the approximations made above to obtain eq.~(\ref{approx_omode}) for the O-mode near fundamental resonance, and linearizing the Boltzmann equation around the Maxwellian distribution function $f_{\rm M}$ with temperature $T_{\rm e}$, the EC quasi-linear diffusion term becomes
\begin{equation}\label{perturbation}
    Q_{\rm EC}(f_{\rm e}) = Q_{\rm EC}(f_{\rm M}) \propto  D \delta(v_\parallel - v_{\parallel,\rm res}) \left({v_\perp^2 \over 2v_t^2}-1\right)
    \exp\left(-{v_\parallel^2 + v_\perp^2 \over 2v_t^2}\right).
\end{equation}
Here the thermal velocity is $v_t = \sqrt{kT_{\rm e}/m_{\rm e}}$.  We thus find that, at the resonant velocity $v_{\parallel,\rm res}$, the waves drive a perturbation characterized by a bulge of electrons at supra-thermal perpendicular velocities and a hole at sub-thermal perpendicular velocities. This wave-driven perturbation carries no current. A net current only arises as a consequence of the subsequent effect of collisions~\cite{fischboozer}. In particular, pitch-angle scattering results in a net transfer of momentum between ions and electrons, because the pitch-angle scattering rate for the positive perturbation at supra-thermal perpendicular velocities is slower than that for the negative perturbation at sub-thermal perpendicular velocities: the hole is filled in more quickly than that the bulge decays. The result is a net current generation. In the final steady state a balance is reached between the EC quasi-linear drive and the collisional dissipation: an electron distribution function is set up, which carries a net parallel current, yet, which transfers no net momentum to ions.

\subsection{A new closure relation for the EC driven current density}

In order to proceed further analytically with the calculation of the Fisch-Boozer current and the derivation of a new closure relation for the evolution of the EC driven current density, we approximate the collisions by a Krook-type collision operator
\begin{equation}\label{krook}
    C(f_{\rm e}) = - \nu(v) (f_{\rm e} - f_M),
\end{equation}
with a velocity dependent collision frequency $\nu(v) = \nu_t (v_t/v)^3$. The momentum loss that is implied by this operator represents the momentum transfer from electrons to ions. In the homogeneous case the solution then is
\begin{equation}\label{homsol}
    \delta f_{\rm e} \equiv f_{\rm e}(v_\parallel, v_\perp; t) - f_M = Q_{\rm EC}(f_M) {1 \over \nu(v)} ( 1 - e^{-\nu(v)t}).
\end{equation}
In most tokamak experiments, however the electron cyclotron wave power deposition is extremely localized along a field line. When we restrict the ECCD power deposition to a finite interval $0 \le x \le L_{\rm EC}$ along a field line and take $t=0$ as the time the power is switched on, the solution to the Boltzmann equation (\ref{boltzmann}) becomes
\begin{equation}\label{solution}
    \delta f_{\rm e} = \begin{cases}
    {Q_{\rm EC}(f_M) \over \nu(v)} \left( 1 - e^{-\nu(v)\min(x/v_{\parallel,\rm res},t)} \right)
    & 0 \le x \le L_{\rm EC} \\
    {Q_{\rm EC}(f_M) \over \nu(v)} \left( 1 - e^{-\nu(v)\min(L_{\rm EC}/v_{\parallel,\rm res}, t - (x-L_{\rm EC})/v_{\parallel})} \right) \times &\\
    \hbox to 3 truecm {}e^{-\nu(v)(x-L_{\rm EC})/v_{\parallel,\rm res}}
    & L_{\rm EC} < x < L_{\rm EC} + t v_{\parallel,\rm res}\\
    0 & x < 0 \hbox{  or  } x \ge L_{\rm EC} + t v_{\parallel,\rm res}  \end{cases}
\end{equation}
The spatial and temporal evolution of the EC driven current density $J_{\rm EC}$ along a magnetic field line can be obtained from straightforward calculation of the first moment of the perturbation $\delta f_{\rm e}$ of the distribution function:
\begin{equation}\label{solution2}
    J_{\rm EC}(x,t) = -e\int{ d^3 v\; v_\parallel \delta f_{\rm e}}.
\end{equation}
The periodicity of the field lines in a tokamak and the long mean free path of electrons means that the electrons will pass through the resonance region multiple times during a collision time. A field line with a non-rational safety factor $q$ fills the flux surface ergodically and thus passes though the resonant region after a multiple of $2 \pi R$ which is on average $L = 4 \pi^2 r R / L_{\rm EC}$.

\begin{figure}
\centering
\includegraphics[width=8cm]{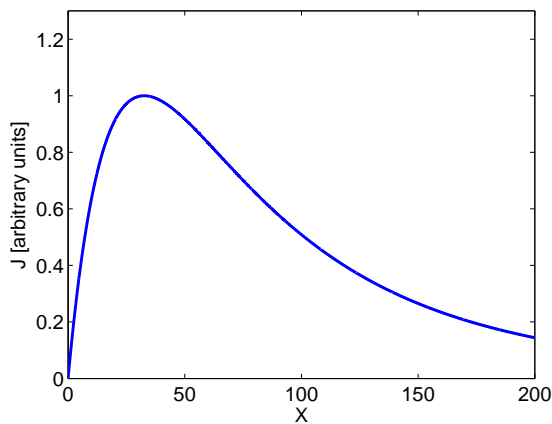}
\caption{The EC driven current density along a field line. The the figure shows the evolution of the EC driven current density according to the  analytical solution of the approximated Fokker-Planck equation given in eq.~\ref{solution}. The length along the field line, $X$ shown on the horizontal axis, is given in normalized units of $v_t/\nu_t$. The EC power is deposited over a narrow region to the left of the figure with $L_{\rm EC} = 10^{-3} \times v_t/\nu_t$. A typical resonant parallel velocity of $v_{\parallel, \rm res} = 2 v_t$ has been used.}
\label{fig.JECCD_along_field_line}
\end{figure}

As an illustration we have calculated the solution for the perturbation $\delta f_{\rm e}$ given by eq.~(\ref{solution}) and the corresponding evolution of the EC current (\ref{solution2}) along a magnetic field line for realistic tokamak parameters. The results are provided in normalized units, with time normalized to the thermal collision time and length to the distance traveled by a thermal electron in a collision time, i.e. $v_t/\nu_t$. For typical tokamak parameters the thermal electron velocity $v_t$ is of order $10^7$~m/s, the thermal collision frequency $\nu_t$ is of order $10^4$ to $10^5$~Hz, and a typical width of the EC power deposition is of order 2 to 10~cm, which corresponds to an ECCD power deposition width of order $10^{-3} \times v_t/\nu_t$. We choose a typical resonant parallel velocity of $v_{\parallel, \rm res} = 2 v_t$. FIG.~\ref{fig.JECCD_along_field_line} shows the EC driven current density established after a large number of collision times ($> 100$), as a function of the length along a field line crossing through the EC power deposition region. What this figure illustrates in a striking manner is that the EC driven current is generated while the perturbation of the distribution function flows out of the EC power deposition region, a tiny region near $x=0$ in FIG.~\ref{fig.JECCD_along_field_line}. Although EC power is deposited in a highly localized region, ECCD generation is a highly nonlocal phenomenon.

We thus obtain the following picture for the process of ECCD: The EC waves create a perturbation in velocity space localized at the resonant parallel velocity, which exists of a velocity space hole at small perpendicular velocities $v_\perp < \sqrt{2} v_t$, and a bulge at high perpendicular velocities $v_\perp > \sqrt{2} v_t$ with a zero net momentum. This perturbation is convected along the field line out of the EC deposition region with the parallel velocity of the resonant electrons. As the perturbation is convected, the velocity space hole at low velocities is filled in more quickly by collisions than the bulge at high velocities is eroded, because of the velocity dependence of the collision frequency. The result is a net current which subsequently decays at the slower collision rate of the high velocity electrons in the bulge.

In a final approximation to arrive at our new closure relation for the EC driven current density, we represent the EC wave driven hole and bulge by two delta functions at perpendicular velocities $v_1$ and $v_2$ with associated collision rates $\nu_i = \nu_t / (v_{\parallel, \rm res}^2 + v_i^2)^{3/2}$, $i=1,2$, so that the perturbation due to the EC waves takes the form:
\begin{equation}\label{two_point_model}
    \delta f_{\rm e} \approx \delta(v_{\parallel} - v_{\parallel, \rm res}) \times \sum_{i=1,2} {c_i\over v_i}\delta(v_{\perp} - v_i).
\end{equation}
Here the amplitudes $c_i$ representing the total numbers of particles missing from the hole and present in the bulge of the velocity space perturbation are driven by the EC waves and decay due to collisions according to
\begin{equation}
    {\partial c_i \over \partial t} =
    \begin{cases} S_i - \nu_i c_i & 0 \le x \le L_{\rm EC}, \\
            -\nu_i c_i & x > L_{\rm EC}.\end{cases}
\end{equation}
The conservation of particles in the EC diffusion leads to a relation of the sources $S_i$ as $S_1 = - S_2$. This is simply the balance between two current density perturbations driven in opposite directions. The current perturbation associated with $v_1$ is driven in the counter direction and the other current perturbation associated with $v_2$ in the co-direction (relative to the net driven current). These two currents can now be represented by the equations
\begin{equation}\label{newmodel1}
    {\partial J_1 \over \partial t} = - S_{\rm EC} - \nu_1 J_1 + v_{\parallel, \rm res} \nabla_\parallel J_1,
\end{equation}
and
\begin{equation}\label{newmodel2}
    {\partial J_2 \over \partial t} = + S_{\rm EC} - \nu_2 J_2 + v_{\parallel, \rm res} \nabla_\parallel J_2,
\end{equation}
where $S_{\rm EC}$ is non zero only between $0 < x < L_{\rm EC}$. The EC driven current density is now defined as the  sum of these two,
\begin{equation}\label{newmodel3}
    J_{\rm EC} \equiv J_1 + J_2.
\end{equation}
Together with a proper definition of the source term $S_{\rm EC}$ this set of equations (\ref{newmodel1}) $-$ (\ref{newmodel3}) represents our new closure model for the EC driven current density. In the limit $\nu_1 \to \infty$, $J_1$ becomes identically zero, and our model reduces to a single-equation model for the evolution of $J_{\rm EC}$ with a single collision frequency $\nu_2$, in the spirit of Giruzzi et al.~\cite{giruzzi1999}. The model proposed by Giruzzi et al. does not include a convective transport term, but relies on a high parallel diffusivity for the equilibration of the driven current density along a field line. In addition, the model of Giruzzi et al. includes a finite perpendicular diffusivity in order to account for the effect of cross-field turbulent transport.

The source term $S_{\rm EC}$ must be related to the EC current drive efficiency, which is defined as the ratio of the total driven current over the absorbed power $\eta_{\rm EC} \equiv I_{\rm EC} / P_{\rm EC}$ under steady state conditions. For the homogeneous case, the steady state solution to equations (\ref{newmodel1}) $-$ (\ref{newmodel3}) is
\begin{equation}
    J_{\rm EC} = S_{\rm EC} \left( {1\over\nu_2} - {1\over\nu_1} \right).
\end{equation}
This implies a relation between $S_{\rm EC}$ and the current drive efficiency $\eta_{\rm EC}$
\begin{equation}\label{source}
    S_{\rm EC} = \eta_{\rm EC} p_{\rm EC} {2 \pi R \nu_2 \over 1 - {\nu_2 \over \nu_1}}
\end{equation}
where $p_{\rm EC}$ is the local (non-flux-surface-averaged) EC absorbed power density. In line with the assumption made at the outset of a linear response of the plasma current to the different driving forces, the EC current drive efficiency is obtained in the standard adjoint approach as~\cite{antonsen1982,karney1986}
\begin{equation}\label{adjoint}
    \eta_{\rm EC} \equiv {\int {\bf S}_w \cdot \nabla_{\bf v} \chi {\rm d}^3v \over \int m_{\rm e} {\bf S}_w \cdot {\bf v} {\rm d}^3v}
\end{equation}
where ${\bf S}_w = -{\bf D}_{\rm EC} \cdot \nabla_{\bf v} f_{\rm e}$ is the wave driven quasi-linear velocity space flux, and $\chi$ is the current response function, which is related to the the solution $f_M \chi$ of the Spitzer-H\"arm equation for the perturbed electron distribution function in the presence of an electric field ${\bf E} = T_{\rm e} {\bf v}_\parallel$~\cite{karney1986}. Generalizations of the response function $\chi$ are available that include toroidal effects like particle trapping as well as a fully relativistic description \cite{cohen1987,linliu2003,marushchenko2008,marushchenko2009}. We do not provide separate equations for the collision frequencies $\nu_1$ and $\nu_2$. In principle these could be obtained from inspection of $Q_{\rm EC}(f_M)$, but are more easily obtained from an analysis of the results of a time dependent bounce-averaged quasi-linear Fokker-Planck calculation~\cite{westerhof1995}. In regimes where the assumption of a linear response breaks down~\cite{harvey1989}, full bounce-averaged quasi-linear Fokker-Planck calculations may also be used to parameterize the nonlinear dependence of the current drive efficiency $\eta_{\rm EC}$ including its predicted synergy with the (orbit averaged) parallel electric field~\cite{james1992,prater2004}.

\section{Model validation}

\begin{figure}
\centering
\includegraphics[width=8cm]{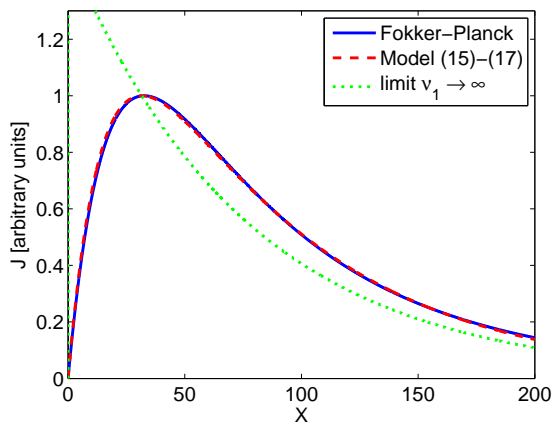}
\caption{Comparison of the EC driven current density along a field line as obtained from the proposed closure model defined in eqs.~(\ref{newmodel1}) $-$ (\ref{newmodel3}) (dashed curve) with the approximated Fokker-Planck solution~(\ref{solution}) (solid curve). The length along the field line is given in normalized units of $v_t/\nu_t$. The plasma and wave parameters are identical to those of FIG.~\ref{fig.JECCD_along_field_line}. The collision frequencies used in the closure model are $\nu_1 = \nu_t/8$ and $\nu_2 = \nu_t/38$, respectively. The figure also shows the result that is obtained from the closure model in the limit of $\nu_1 \to \infty$ (dotted curve).}
\label{fig.JECCD_compare}
\end{figure}

In a first step of validation of our closure model for the EC driven current density, we demonstrate that the evolution of the EC driven current density along a field line as predicted by the equations (\ref{newmodel1}) $-$ (\ref{newmodel3}) captures the behavior of this driven current density as predicted by the solution given in eq.~(\ref{solution}) of the approximated quasi-linear Fokker-Planck equation and displayed in FIG.~\ref{fig.JECCD_along_field_line} with high accuracy. In particular, for the perpendicular velocity of the velocity space hole we chose $v_1 = 0$, and for the bulge we chose $v_2 = 2.7 v_t$ close to the maximum of $Q_{\rm EC}(f_M)$ (\ref{perturbation}). With the resonant parallel velocity $v_\parallel = 2 v_t$ as used in the example, this corresponds to collision frequencies $\nu_1 = \nu_t/8$ and $\nu_2 = \nu_t/38$, respectively. The result shown in FIG.~\ref{fig.JECCD_compare} by the dashed curve is an almost perfect match between the closure model and the solution of the approximated quasi-linear Fokker-Planck equation (plotted as a solid line). In the same figure, we also show the result that is obtained in the limit of $\nu_1 \to \infty$ holding $\nu_2$ constant (dotted curve), corresponding to a single equation model similar to the anisotropic diffusion model of Giruzzi et al.~\cite{giruzzi1999}. Note that the very strong localization of the EC wave power results in a large overestimate of the local current density in the power deposition region in this limit of $\nu_1 \to \infty$.

\begin{figure}
\centering
\includegraphics[width=8cm]{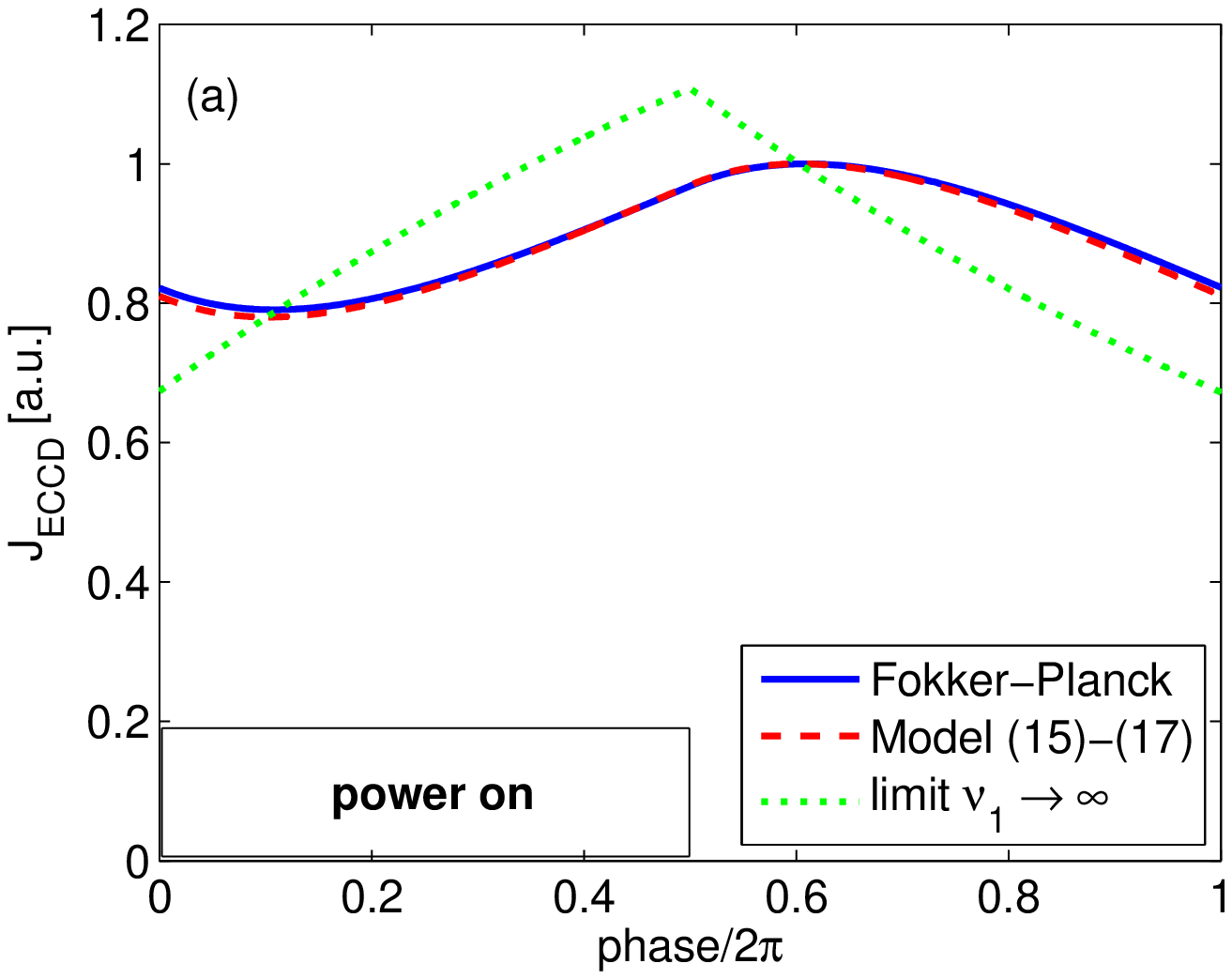}\includegraphics[width=8cm]{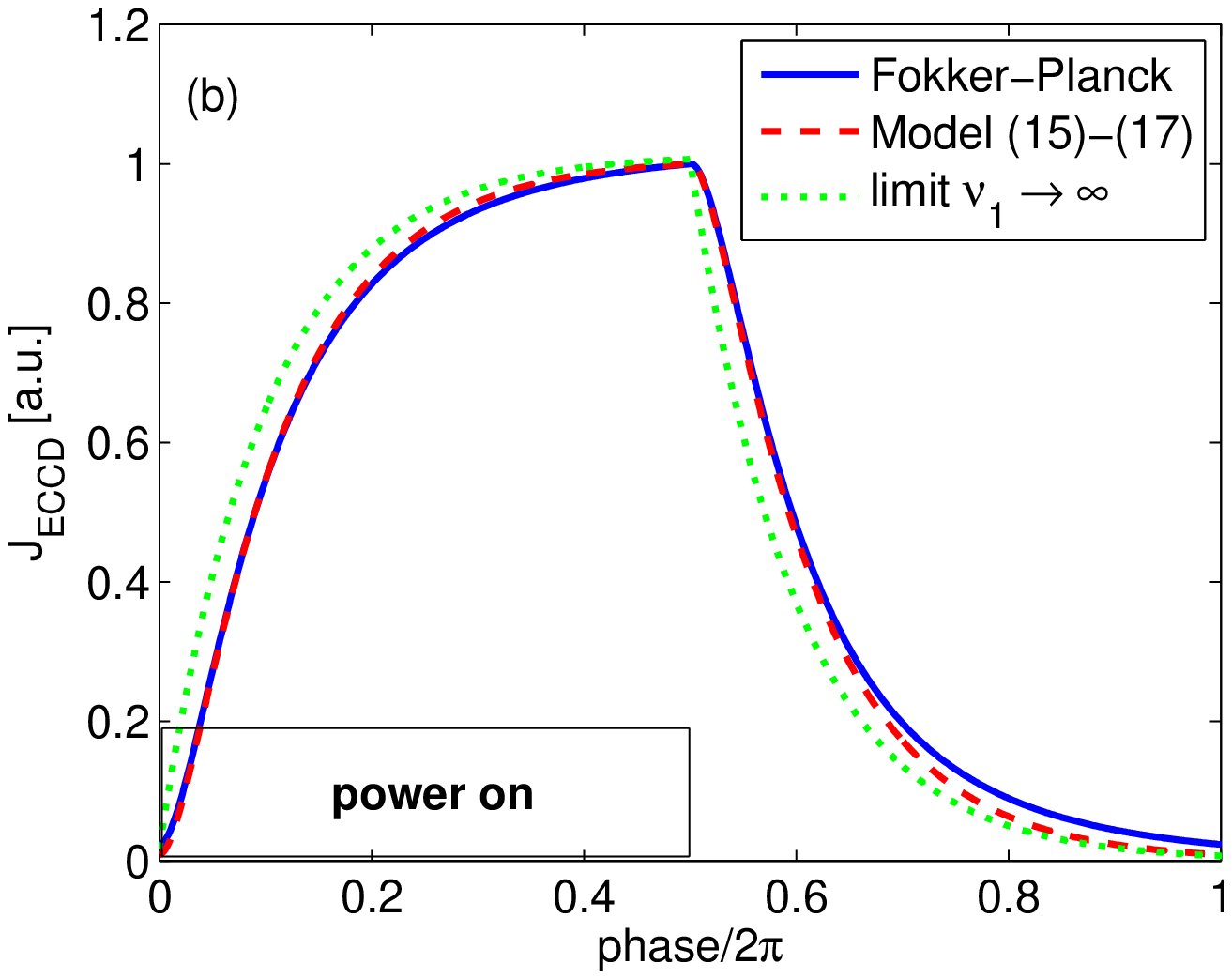}
\caption{Modulated ECCD. The figure shows the results of calculations for the evolution of the driven current density during one period of EC wave power modulation. Homogeneous wave power deposition on the flux surface is assumed. The plasma, wave, and model parameters are identical to those used in FIGs.~\ref{fig.JECCD_along_field_line} and \ref{fig.JECCD_compare}. A 50\% duty cycle of the wave power modulation is assumed with a modulation period of (a) $38 \times \nu_t^{-1}$ which is equal to the collision time of the fast electrons in the bulge of the distribution function created by the ECCD and (b) $380 \times \nu_t^{-1}$. Solid curves show the results of the solution~(\ref{solution}) to the approximated quasi-linear Fokker-Planck equation, dashed curves the results from the proposed closure model defined in eqs.~(\ref{newmodel1}) $-$ (\ref{newmodel3}), and dotted curves the results of the proposed model in the limit of $\nu_1 \to \infty$.}
\label{fig.JECCD_modulation}
\end{figure}

In a tokamak, a field line covers the flux surface ergodically or closes upon itself in case of rational values of the safety factor $q$. Along the field line it will pass many times through the power deposition region, with the currents from each of these passages adding up. For a highly localized EC power source the average distance between subsequent passages through the power deposition region is of order $4\pi^2 r R / L_{\rm EC}$ and the average time between subsequent passages is of the order of a thermal collision time. As a result of the fast convective transport, the driven current density is almost constant on a flux surface and is reasonably approximated by the assumption of a homogenous spread of the wave power over the flux surface. We use this assumption to analyze the temporal dynamics of the EC driven current density in case of modulated ECCD. Modulation of the ECCD power is applied to improve the stabilizing effect of the EC driven current density on NTMs. Moreover, rotation of a magnetic island results in a natural modulation of the flux-surface averaged power deposition even in case of continuous ECCD. FIG.~\ref{fig.JECCD_modulation} shows the temporal evolution of the EC driven current density during a modulation period. The parameters are identical to those of the previous sets of calculations except that we now solve for the homogeneous case, so that the convective term drops out of the equations. The full curve again shows the results of the approximated quasi-linear Fokker-Planck equation and the dashed curve the results obtained with the closure model for the EC driven current density defined in eqs.~(\ref{newmodel1}) $-$ (\ref{newmodel3}) using the same collision frequencies as before: $\nu_1 = \nu_t/8$ and $\nu_2 = \nu_t/38$. Two cases are shown: one with a modulation period equal to the collision time of the fast particles in the bulge, i.e.~$T_{\rm modulation} = 38 \nu_t^{-1}$ (a), and one with a much slower rotation period $T_{\rm modulation} = 380 \nu_t^{-1}$~(b). In both cases the duty cycle of the ECCD power modulation is 50\%. A good agreement is obtained between the results from the approximated quasi-linear Fokker-Planck and the closure model in all cases. We again compare also with the model in the single equation limit $\nu_1 \to \infty$ indicated by the dotted lines in the figures. The latter model clearly does not give a good representation of the dynamical evolution of the EC driven current density, in particular, in the case of fast power modulation.

\subsection{Validation on full bounce-averaged quasi-linear Fokker-Planck}

In a recent paper, Ayten et al.~\cite{ayten2014} studied the dynamical evolution of the EC driven current density inside a rotating magnetic island by means of full bounce-averaged quasi-linear Fokker-Planck calculations. A case study was made for parameters that are representative of experiments on $m=3$, $n=2$ NTM suppression by ECCD in ASDEX-Upgrade~\cite{reich2012}. In this subsection we show that our closure model also provides a good approximation to the dynamics of the EC driven current density as obtained from these complete Fokker-Planck calculations. Full details of the discharge and plasma parameters can be found in~\cite{ayten2014}. Here we repeat the most relevant parameters. The major radius of the magnetic axis in this discharge is at $R=1.70$~m, and the position of the resonant $q=3/2$ surface is at a normalized minor radius $x=0.4$. The plasma density and temperature at $q=3/2$ are $n_{\rm e} = 6.56 \times 10^{19}$~m$^{-3}$ and $T_{\rm e} = 2.7$~keV. An effective charge of the plasma ions of $Z_{\rm eff} = 1.6$ was used in the calculations. This yields a thermal electron collision time of $\tau_{\rm e} = 3 \times 10^{-5}$~s. In line with the bounce averaging of the Fokker-Planck equation, we apply our model in the homogeneous limit assuming that the parallel convection is sufficiently fast that the EC driven velocity space distribution and current density are effectively constant over a flux surface. We thus focus on the temporal evolution of the driven current density during one island rotation period. The relevant collision frequencies $\nu_1$ and $\nu_2$ in our closure model equations (\ref{newmodel1}) $-$ (\ref{newmodel3}) are estimated by fitting the rise of the current density to steady state on an equilibrium flux surface in the middle of the power deposition profile. In this case, we obtain $\nu_1 = 88.4$~kHz and $\nu_2 = 15.2$~kHz. These numbers are consistent with the only mildly super-thermal character of the ECCD current-carrying electrons in this case with a toroidal injection angle relatively close to perpendicular $\phi_{\rm inj} = -8^{\rm o}$. The EC wave is injected so that it propagates nearly tangentially to the flux surfaces in the region of power deposition. Therefore, these collision frequencies are averages over the absorbed power profile. Ayten et al. note good agreement between adjoint and full bounce-averaged quasi-linear Fokker-Planck calculations of the current drive efficiency. In our calculations we will be using an efficiency of $\eta_{\rm EC} = -0.0085$~A/W, obtained from the adjoint calculation (\ref{adjoint}) implemented in the TORAY ray-tracing code~\cite{toray1,toray2,linliu2003,marushchenko2008,marushchenko2009}. The negative value of $\eta_{\rm EC}$ in this case corresponds to co-current drive in the direction of the plasma current (i.e. clock wise in the direction of negative toroidal angles). The power densities as a function of time at the different flux surfaces are taken from the Fokker-Planck calculations and were consistent with ray-tracing calculations~\cite{ayten2014}.

\begin{figure}
\centering
\includegraphics[width=8cm]{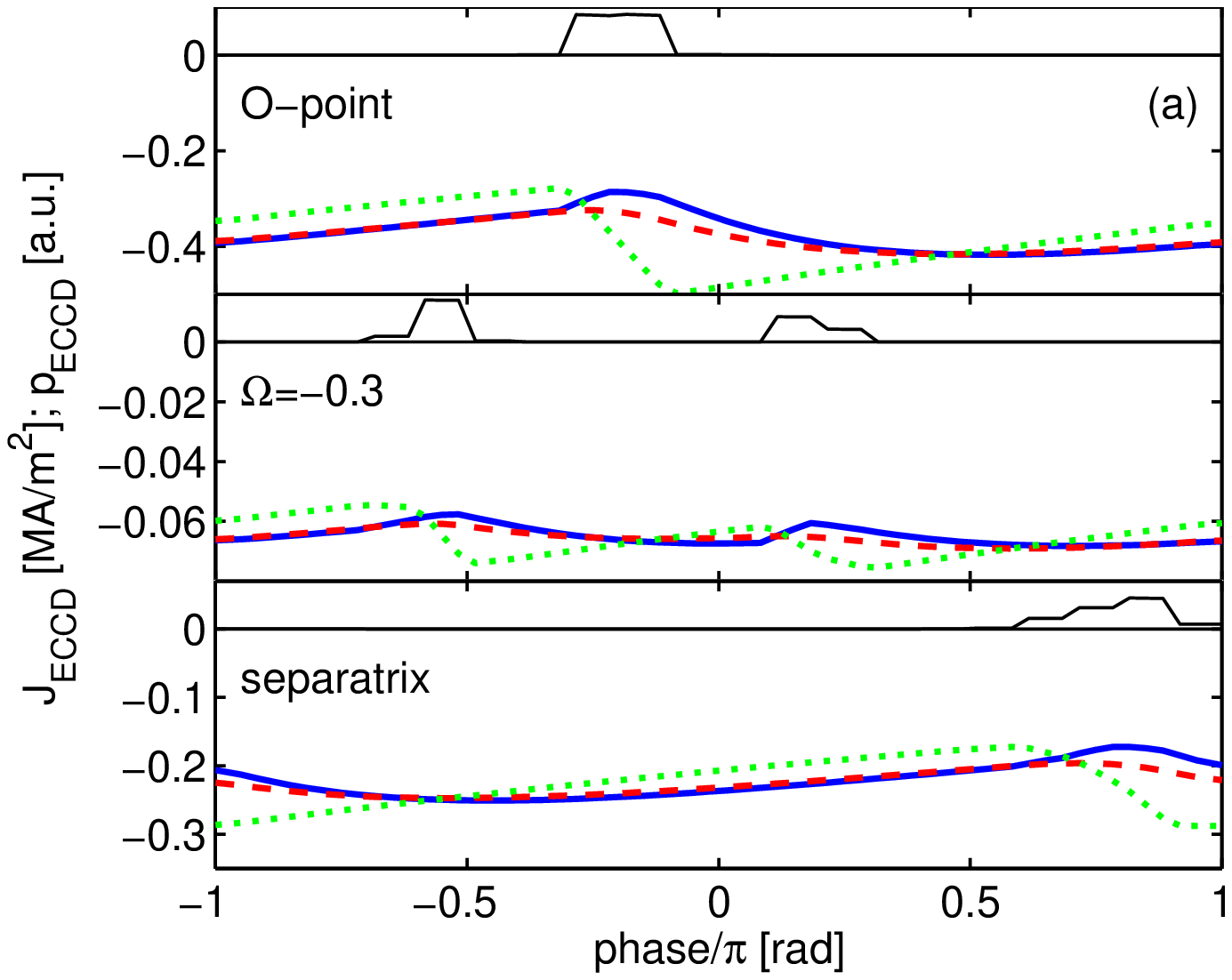}\includegraphics[width=8cm]{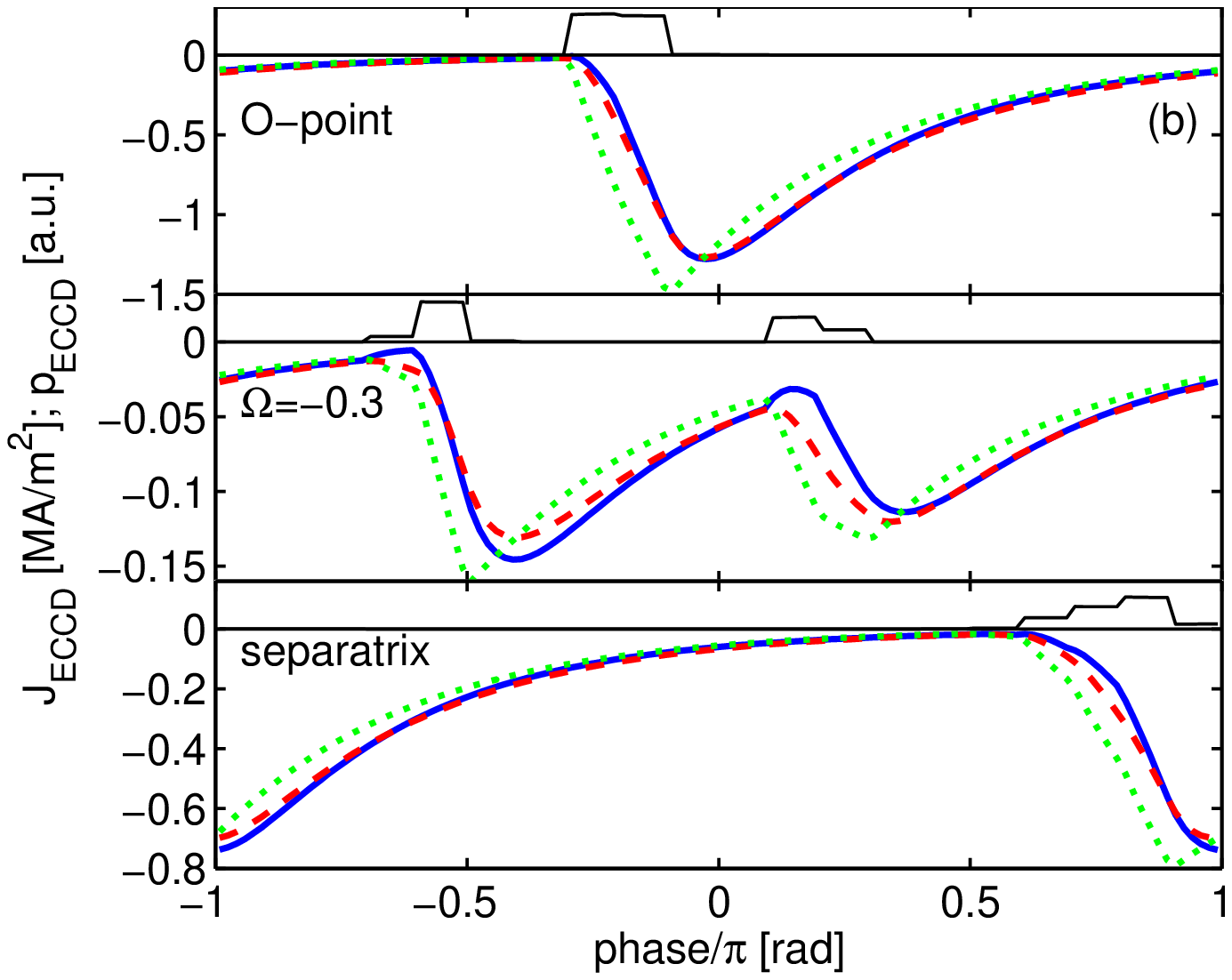}
\caption{ECCD in a rotating magnetic island. The figure shows the evolution of the EC driven current density during one rotation period of the magnetic island. The current density is shown at three representative locations: near the O-point, about midway between the O-point and the separatrix at a normalized island flux coordinate~$\Omega = -0.3$, and immediately outside the separatrix. Solid curves show the results of full bounce-averaged quasi-linear Fokker-Planck simulations~\cite{ayten2014}, dashed curves the results of our proposed closure model (\ref{newmodel1}) $-$ (\ref{newmodel3}) with the current drive efficiency obtained from the adjoint calculation implemented in the ray-tracing code, and dotted curves again the results of the proposed model in the limit of $\nu_1 \to \infty$. The thin solid lines indicate the times during a rotation period that the surface is heated by showing the (positive) EC power density in arbitrary units.}
\label{fig.rotating_island}
\end{figure}

The Fokker-Planck calculations of~\cite{ayten2014} have been obtained with the RELAX bounce-averaged quasi-linear code~\cite{relax}, which has been extended to include the averaging over the perturbed flux surfaces inside a magnetic island. We will compare the predictions from the closure model equations (\ref{newmodel1}) $-$ (\ref{newmodel3}) and a source determined by eq.~(\ref{source}) with the evolution of the EC driven current density from the Fokker-Planck simulations in the case of a rotating magnetic island on a number of representative flux surfaces: near the O-point of the magnetic island, at a surface midway in the island, and at the separatrix. The power deposition region is much smaller than the 8~cm wide magnetic island (full distance between separatrices measured on the low field side). Two simulations are performed, one corresponding to an island rotation frequency of 23~kHz consistent with the experimentally observed mode rotation, and another one with a significantly smaller rotation frequency of 3~kHz. In the first simulation, the rotation frequency is of the same order as the collision frequency of the EC current-carrying electrons ($\nu_2 = 15.2$~kHz), while in the second simulation the rotation frequency is significantly lower. The results are given in FIG.~\ref{fig.rotating_island}. The full curves correspond to the results of the full bounce-averaged quasi-linear Fokker-Planck calculations and are taken from Figure~11 of Ayten et al.~\cite{ayten2014}. The dashed curves represent the results obtained with our closure for the EC driven current density, and the dotted curves the results obtained in the single equation limit $\nu_1 \to \infty$. Our closure model for the EC driven current density clearly provides a good representation of the dynamic evolution of the EC driven current density inside a rotating magnetic island. In particular, it provides an accurate model for the delayed response of the driven current evolution in reaction to the EC power deposition. In the single equation limit $\nu_1 \to \infty$, similar to the model proposed by Giruzzi et al.~\cite{giruzzi1999}, the immediate response of the driven current to the EC power density clearly gives a much less accurate representation of the time evolution of $J_{\rm EC}$.

In the case of the 23~kHz high frequency rotation, an immediate and opposite response in the current density obtained from the full bounce-averaged quasi-linear Fokker-Planck simulations is observed when the flux surface moves through the power deposition region. This is a consequence of the trapping of resonant electrons, which results in an immediate generation of an oppositely directed current: the so-called Ohkawa current~\cite{ohkawa}. The Ohkawa current is not described by our closure model which is designed to model the Fisch-Boozer current drive. However, the Ohkawa effect is included in the adjoint calculation and the full bounce-averaged quasi-linear Fokker-Planck modeling of the EC current drive efficiency. For the parameters considered here, the Fisch-Boozer mechanism is the dominant current drive effect with the Ohkawa current forming a small correction. In cases where the Ohkawa current dominates, the present closure model could provide a reasonable simulation by choosing $\nu_1 = \infty$.

\section{Summary and discussion}

The main result of this paper is formed by equations (\ref{newmodel1}) - (\ref{newmodel3}) and (\ref{source}), which provide a new closure model for the EC driven current density that faithfully represents the nonlocal character of the driven current and its origin in the Fisch-Boozer mechanism \cite{fischboozer}. In single fluid MHD modelling, this closure model is to be used in combination with the usual modification of Ohm's law (\ref{ohm}). In the model, the EC driven current density is the sum of two contributions representing the EC driven quasi-linear modification of the electron distributing function: a hole at low perpendicular velocities and a bulge at high perpendicular velocities. The dominant transport is provided by parallel convection of these perturbations with the parallel resonant velocity. This parallel convection is the main transport mechanism that will result in almost constant (in space) driven current density over closed magnetic surfaces.

The new closure model for the EC driven current density has been validated on the analytical solution of the approximated Fokker-Planck equation discussed in Section~2. This validation addressed the strongly non-local character of the evolution of the EC driven current density along a magnetic field line, which is a result of the fast parallel motion of the electrons responsible for the EC driven current density. In practice this leads to almost constant (in space) driven current density over closed magnetic surfaces, an assumption that lies at the basis of the usual bounce-averaged quasi-linear Fokker-Planck code modelling of ECCD. In the implied homogenous limit of EC power deposition, the dynamic evolution of the driven current density in response to a modulated power source has been calculated. The proposed closure model is shown to capture with great accuracy the dynamical evolution as calculated with the approximated quasi-linear Fokker-Planck equation.

A further validation of the new closure model for the EC driven current density has been provided by comparison of flux surface averaged predictions of the model with full bounce-averaged quasi-linear Fokker-Planck code simulations of ECCD inside rotating magnetic islands. The Fokker-Planck code results used in this comparison, were obtained with the RELAX code and are taken from a recent paper by B. Ayten et al. on the Fokker-Planck code modelling of ECCD for NTM suppression in ASDEX Upgrade~\cite{ayten2014}. The closure model is shown to give a very good representation of the driven current evolution. The only exception is formed by a small immediate response of the driven current density to the EC power deposition as a consequence of the Ohkawa effect due to EC wave induced particle trapping~\cite{ohkawa}.

In the limit that the collisionality of the electrons in the hole goes to infinity, i.e. $\nu_1 \to \infty$ in eq.~(\ref{newmodel1}), the new model reduces to a single-equation closure, similar to the model proposed by Giruzzi et al.~\cite{giruzzi1999}. The anisotropic diffusive model of Giruzzi et al. does not include the parallel convection, which in our proposed model is the dominant mechanism responsible for fast equilibration of the driven current density over a closed flux surface. Instead, the model of Giruzzi et al. relies on a high parallel diffusivity for the effective flux surface averaging. In addition it includes a perpendicular diffusivity to account for the cross-field transport as a consequence of plasma turbulence. In case of narrow EC power deposition, this cross-field transport becomes important~\cite{bertelli2009} and can be added in a trivial manner in our model. The direct response of the EC driven current to the EC power deposition as implied by the limit $\nu_1 \to \infty$ does not correctly capture the dynamic evolution of the EC driven current density in case of fast modulation of the EC wave power. Such a fast modulation of the wave power occurs naturally on the magnetic surfaces inside rotating magnetic islands, when ECCD is applied for NTM stabilization.

Finally, we note that in~\cite{jenkins2010} an alternate approach is followed to simulate the effect of ECCD on tearing modes in a nonlinear 3D MHD simulation: instead of accounting for the modification of the resistivity, the (sub-dominant) localized RF parallel force as introduced in~\cite{hegna2009} is included in Ohm's law. In this case, compressional Alfv\'en waves are found to play a crucial role in the equilibration of the current density along magnetic field lines. In the case of ECCD, this method neglects the dominant Fisch-Boozer current which is the result of the quasi-linear diffusion in the perpendicular velocity direction. The Fisch-Boozer current is correctly modeled by the closure model derived in this paper.

\section*{Acknowledgment}
\noindent This work has been performed in the framework of the NWO-RFBR Centre of Excellence (grant 047.018.002) on Fusion Physics and Technology. This project has received funding from the European Union's Horizon 2020 research and innovation programme under grant agreement number 633053. The views and opinions expressed herein do not necessarily reflect those of the European Commission. We acknowledge stimulating discussions with C.C.~Hegna during the Lorentz Center workshop Modeling Kinetic Aspects of Global MHD Modes, 2-6 December 2013, Leiden, the Netherlands.

\end{document}